\documentclass[sigconf, screen]{acmart}
\usepackage{amsfonts}
\usepackage{multirow,tabularx}
\usepackage{xcolor,colortbl}

\definecolor{Gray}{gray}{0.85}
\definecolor{LightGray}{gray}{0.95}
\newcolumntype{a}{>{\columncolor{Gray}}c}

\copyrightyear{2023}
\acmYear{2023}
\setcopyright{licensedusgovmixed}\acmConference[CSET 2023]{2023 Cyber Security Experimentation and Test Workshop}{August 7--8, 2023}{Marina del Rey, CA, USA}
\acmBooktitle{2023 Cyber Security Experimentation and Test Workshop (CSET 2023), August 7--8, 2023, Marina del Rey, CA, USA}
\acmPrice{15.00}
\acmDOI{10.1145/3607505.3607524}
\acmISBN{979-8-4007-0788-9/23/08}

\begin{document}
\title{Battle Ground: Data Collection and Labeling of CTF Games to Understand Human Cyber Operators}

\author{Georgel M. Savin}
\affiliation{%
  \institution{United States Naval Academy}
  \streetaddress{1 Wilson Road}
  \city{Annapolis}
  \state{Maryland}
  \country{USA}
}

\author{Ammar Asseri}
\affiliation{%
  \institution{United States Naval Academy}
  \streetaddress{1 Wilson Road}
  \city{Annapolis}
  \state{Maryland}
  \country{USA}
}

\author{Josiah Dykstra}
\affiliation{%
  \institution{National Security Agency}
  \city{Ft. George G. Meade}
   \state{Maryland}
  \country{USA}
}

\author{Jonathan Goohs}
\affiliation{%
 \institution{Cyber Strike Activity Sixty-Three}
  \city{United States Navy}
   \country{USA}
  }

\author{Anthony Melaragno}
\affiliation{%
  \institution{United States Naval Academy}
  \streetaddress{1 Wilson Road}
  \city{Annapolis}
  \state{Maryland}
  \country{USA}}
  
\author{William Casey}
\affiliation{%
  \institution{United States Naval Academy}
  \streetaddress{1 Wilson Road}
  \city{Annapolis}
  \state{Maryland}
  \country{USA}}

\begin{abstract}
Industry standard frameworks are now widespread for labeling the high-level stages and granular actions of attacker and defender behavior in cyberspace. While these labels are used for atomic actions, and to some extent for sequences of actions, there remains a need for labeled data from realistic full-scale attacks. This data is valuable for better understanding human actors' decisions, behaviors, and individual attributes. The analysis could lead to more effective attribution and disruption of attackers.

We present a methodological approach and exploratory case study for systematically analyzing human behavior during a cyber offense/defense capture-the-flag (CTF) game. We describe the data collection and analysis to derive a metric called keystroke accuracy.  After collecting players' commands, we label them using the MITRE ATT\&CK framework using a new tool called Pathfinder. We present results from preliminary analysis of participants' keystroke accuracy and its relation to score outcome in CTF games.  
We describe frequency of action classification within the MITRE ATT\&CK framework and discuss some of the mathematical trends suggested by our observations.  
We conclude with a discussion of extensions for the methodology, including performance evaluation during games and the potential use of this methodology for training artificial intelligence.
  
\end{abstract}

\ccsdesc[500]{Security and privacy~Network security}
\ccsdesc[300]{Security and privacy~Human and societal aspects of security and privacy}

\keywords{capture the flag, game theory, penetration testing, ATT\&CK}

\maketitle

\section{Introduction}
Cyber operational skills and education are important, but assessment techniques are lacking.  In short, one of the critical questions is how can we measure technical skills and enable learning to match the needs of current cyberspace operational environments?  Within the cyber industry, there is a wide range of skills that people can demonstrate.  For example, someone who excels at binary exploitation is in no way guaranteed to show a promising skill set at social engineering.  This wide range of skills that the cyber industry requires also presents a large landscape to be evaluated and measured for effectiveness and performance. This paper asserts how players within simulated cyber space games leverage critical thinking and technical knowledge to solve problems with an offensive cyber operational mindset.

To address the critical questions of how operator actions translates to outcomes, we design human experiments where measureable results can be correlated with action decisions.  This is done by using a Hack the Box (HTB) challenge and analyzing the differential outcome scores in relation to differential human subject input action sequences.  This analysis is essential to the industry, as understanding cyber operator's methodology on attempting to gain a foothold on a device and network provides great insight on how information security and cyber defense professionals can advance their tactics, techniques, and procedures to protect their assets and organizations at a larger scale and higher success rate.

Platforms such as HTB provide users with the opportunity to practice cyber defense and penetration testing within a safe sandbox environment~\cite{htb}. These training platforms also offer competitive games, such as capture-the-flag (CTFs) events where different cyber techniques can be applied within the HTB virtual network. CTFs are used to test operational skills, build teams and develop teamwork, and advance organizational cybersecurity posture by developing a skilled workforce of ethical hackers, penetration testers, and network defenders.  The platform and competitive nature encourages those to collaborate and fuse their skills, as many elements of cyber space come into play on virtualized boxes.  It is important to highlight the HTB layout for this paper analyzed in what we would describe as a basic attack platform for cyber operators. 

In cyberspace, an exploit is a piece of code or program used to affect a system based on a weakness in that very system, which is known as a vulnerability. Defensive cyberspace actions and information security efforts respond to malicious cyber actions by attackers through conducting incident response and patching vulnerabilities to systems.  These patches can be pushed on an automated timeline, or as a standalone patch in response to a reported incident that poses a serious enough threat to a network to warrant immediate action. The HTB players aim an exploit against a specific vulnerability they have identified on the simulated ``box.''

The platform also encourages the sharing of tactics, techniques, and procedures in social media settings, contributing to the development of the student and team, if applicable for a competitive setting. Due to its realistic nature, HTB can even be used for recruiters to assess an applicant's skill set when applying to cyber related jobs. These systems offer practical instances of real cybersecurity problems and increasingly mirror the essential challenges at play in real world cybersecurity problems. 

The paper is organized as follows. Section~\ref{sec:relatedWork} describes related work. In Section~\ref{sec:instrumentation}, we present our approach to instrumentation and data collection and then labeling and analysis in Section~\ref{sec:labelingAndAnalysis}. Section~\ref{sec:CaseStudy} contains a preliminary case study, analysis, and findings. We conclude in Section~\ref{sec:Conclusion}.

\section{Related Work}
\label{sec:relatedWork}
Capture-the-flag and other cyber competitions are regularly used to educate and train people in cybersecurity. CTF games commonly offer the ability to collect game-related data, from keystrokes and screenshots to packet captures~\cite{taylor2017ctf}. This data may be substituted in research as a proxy for real malicious traffic. For instance, the DEF CON data sets have been widely used in research and included packet captures~\cite{defconCTFarchive}. Data sets from collegiate competitions have included packet captures, logs, and other data~\cite{munaiah2019cybersecurity}. These data offer many potential applications, including game theory~\cite{goohs2022reducing}.
In 2021, {\v{S}}v{\'a}bensk{\`y} et al. collected and released a dataset of 13,446 commands from 175 participants in cybersecurity training events~\cite{vsvabensky2021dataset}. While the events lacked time pressure associated with CTF, the labeled dataset includes command-level timing information but not individual keystrokes.

To a lesser extent, CTF data sets have also been used for research to understand the players. 
Bashir et al. used a survey to profile cybersecurity competition participants, such as their tendency to be high in openness and investigative interests~\cite{bashir2017profiling}. In a study of tactical cyber operators performing their real work, Dykstra and Paul found evidence that fatigue and cognitive workload impact operations, but the study did not collect keystrokes as a means of detecting stress~\cite{paul2017understanding}. Most of this prior work has not considered the command-level behavior of individual players and the relationship to player attributes. 

Some researchers have pointed out that traditional CTF events were not designed for measuring human performance and that controlled red team experiments can be advantageous~\cite{ferguson2019world}. For example, Johnson et al. used lab-based experiments to study deception and decision making~\cite{johnson2021decision}. Our work seeks to overcome some of these limitations by creating new data collection and analysis of human behavior that would work equally well in controlled experiments.

There has also been research to explore the design, scoring, and data collection of CTF games~\cite{davis2014fun}. More broadly, cyber ranges and testbeds may also be instrumented for data collection from experimentation and used to host competitions~\cite{yamin2020cyber}. 

As security professionals have become more interested in detecting sophisticated attacks, there is growing interest in understanding and representing action sequences over singular indicators, such as hashes of malicious files. Most closely related to our work is a 2022 evaluation of the vulnerability assessment methodologies among players in a collegiate competition ~\cite{meyers2022examining}. The researchers constructed timelines of participant behavior and labeled them using MITRE ATT\&CK framework. Their goal was to understand how people actually behave in vulnerabilities discovery, and it did not consider collection or analysis of keystroke timing or accuracy.
In 2022, the MITRE Engenuity Center for Threat-Informed Defense also created Attack Flow, a language for describing the sequence of techniques used by an attacker ~\cite{mitre2023flow}. This offers one way to document and visualize behavior.

CTF studies often focus on offensive and attack behavior as a means of improving defenses and defenders. To our knowledge, there are no published studies labeling the command-line behavior of defenders. Such data could prove valuable in evaluating defenders and in creating tools and procedures that help them overcome areas of weakness.

Finally, we take inspiration from the field of keystroke dynamics. A body of research has emerged using the timing intervals between key presses as a away to authenticate and identify individual users~\cite{tsimperidis2021age}. This technique has many applications, including identification of individual programmers~\cite{longi2015identification}. We decided to collect keystroke data for the opportunity to mimic attacker behavior at a granular level, including typing accuracy and speed.

\section{Instrumentation and Data Collection}
\label{sec:instrumentation}
To create a baseline understanding of how players play CTF games, we requested human subjects familiar with CTF games to enter a human study where the game play could be monitored.
To detail our collection methodology we provide a summary of software requirements and implementation for the three main stages. 
\begin{itemize}
    \item Stage 1: The design and development of software (and its integration of existing software components) facilitating the capture of user inputs with high resolution.   
    \item Stage 2: Control and validation scripts to ensure that data collection within the deployed environments is online and collecting data.
    \item Stage 3: Data normalization routines which assemble commands from sequences of sub command data entry and calculate accuracy.
\end{itemize}

\subsection{Stage 1: Collection Screen}  The requirement of the first stage are host-based, and aimed to collect user actions from the virtual host that each player initializes actions from in the remote web-based CTF game.  Note many of the same techniques could be applied (without any changes) to local or offline CTF games as well.  
Note that in CTF games, a user may operate from multiple systems, we therefore use the term {\it screen} to denote the player's primary system, or the system which accepts user input first, before it is transmitted to other systems and remotely executed.  
The only essential requirement is access to modify the software environment (or operating system) of the screen system which users use as their primary base of operation 
the desired collection at the screen include:

\begin{enumerate}
    \item \textbf{Command History}. In order to capture all the commands that were run locally on the command line, we use the native \texttt{history} feature in the bash shell. We configured bash to record timestamps with each command. 
    
\item \textbf{Keystrokes}. We use the \texttt{logkeys} keylogger to capture the keystrokes of the commands that are being used remotely (e.g., the user has connected to a vulnerable box and proceeds to exploit it)~\cite{logkeys}. 

\item \textbf{Screen Recording}. In order to analyze other interactions, such as browser interactions and web searches, we captured a complete video recording using \texttt{recordmyscreen}~\cite{recordmyscreen}.
\end{enumerate}

{\bf Implementation. } For the keystroke monitor and the screen recording, we use the following Linux open source software projects:  
logkeys and recordmydesktop.  Our installation scripts are available at \url{ https://github.com/xmohemx/HSM- }. 

\subsection{Stage 2: Control, Validation, and Aggregation}
Instrumenting a single screen involves ensuring collection components are running during the user interactions.
However, instrumenting multiple screens on a set of hosts entails additional challenges.  
To address these challenges we designed control software with capabilities to remote deploy component, start/stop collection, verification of collection status, and provide recover options should a player reconfigure a keyboard during the game session.  These and additional tasks of data log tracking and marshaling to a central repository are subsumed by the second stage, what we refer to as control software.  

{\bf Implementation. } 
We create a command and control server in which we are able to see in real time the flow of the captured data. In order to realise this we use the Linux tool Cluster SSH~\cite{clusterssh}. Cluster SSH allows us to connect to multiple interfaces at the same time and supervise the files where the keystrokes are be saved. In order to watch the flow of the collected data we use the \texttt{tail} command~\cite{tail}. 

After the session, we export the scoreboard and all the interactions with the vulnerable system in a JSON format.

To make sure that there is anonymity and uniformity across all devices, we developed a script that encompasses all the needed material for our research. We used this model to help us install all the needed programs to collect the data, create a specific folder in the home directory for the collected data, and enabled ssh to create a hub of all connected computers. This hub is where the data can be extracted from a main central computer using clusterssh to ensure data integrity is not compromised and make the gathering piece of data more mainstream. The users will be able to login into a machine and use it without any disturbance as we made sure that the scripts included and the programs used to record the research were both lightweight and effective in gathering all the necessary information to achieve.
Finally, logs on the individual hosts are collected using Cluster SSH and aggregated to the command server. 

\subsection{Stage 3: Keystroke Normalization} 
Raw keystroke collection is collected at the granularity of each key press event, and as such must undergo additional processing to re-synthesize the individual actions described by command entry (commands with options and arguments).  Individual commands are likely to be the atomic actions in which players assemble for strategic ends in the game.  Hacker challenges often teach these individual commands (with options and arguments) as part of training modules leading up to game competitions.  

\begin{figure}[h!]
    \centering
    \includegraphics[width=80mm]{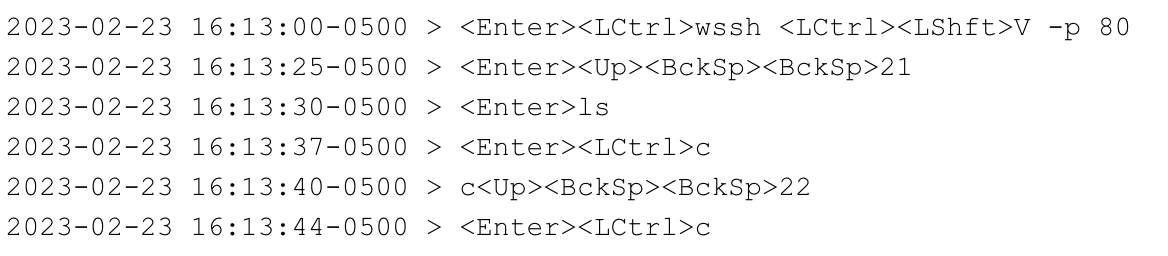}
    \caption{Keylogger output offers fine grain view of each keypress event.  To interpret data rules consistent with the operator's terminal shell must be applied to normalize data to command intentions of the human player.}
    \label{fig:keylog}
\end{figure}

{\bf Normalization algorithm.} To recover command actions from keystrokes, we interpret the keystroke data similarly to the Bash shell (often the default command shell on Linux systems).  An example of command entry is shown in Figure~\ref{fig:keylog}.  Note that the shell provides a modifiable string buffer submitted upon a submit event (usually control key <enter>), and there are two essentially different types of key press actions, those which emit a character to the buffer and those which control buffer editing or entry such as enter, backspace, delete, left, right, control-C and so forth.  Since many of the control characters can modify the string buffer, we make sure to recapture the intended command.  For completeness, we provide our command reconstruction which assumes a character array ${\sc BUFF} $ and counters $k, K$ which track the cursor position and the right extent of the inserted string, and apply the following algorithm and rules given input $x$.  The function ${\sc REC}$ will record the normalized entered command.
\begin{equation*}
    F  = \begin{cases}
        {\sc REC}(BUFF[0:K]), k = 0, K = 0 \text{ if } x = <enter> \\
        k = 0  \text{ if } x = <ctl> a \\
        k = K \text{ if } x = <ctl> e \\
        k = min( k + 1, K ) \text{ if } x = <right> \\
        k = max( 0, k - 1) \text{ if } x = <left> \\
        k = min( k + 1, K ) \text{ if } x = <ctl> F \\
        k = max( 0, k - 1) \text{ if } x = <ctl> B\\
        k = max( 0, k - 1) , K = max( 0, K-1) \text{ if } x = <BckSp> \\
        BUFF[k:K] = x BUFF[k:K], k++, K++ \text{ otherwise  }. 
    \end{cases} 
\end{equation*}
The rules above account for many of the control data entry observed in the experiment, however, we did not implement command line history (<up>) which requires identification of terminal shell session.  However, the rules above facilitate reconstruction of many commands which would otherwise be garbled due to cursor position movement.

\subsection{Keystroke Accuracy}  
\label{sec:keystrokeAccuracy}
With the data normalization method defined, we consider a measure for {\it keystroke accuracy} as the ratio of length of the final submitted command buffer (i.e., $K$) to the number of total keystroke events.
Letting $C$ be a counter set to zero when a command is submitted (i.e., <enter> is pressed with a non-empty command buffer detected with condition $K>0$ ). 
The counter $C$ will be incremented for each keystroke, in order to count the total number of keys pressed till the next submission event.  We define the keystroke accuracy as: 
\begin{equation*}
    A = \frac{ K }{ C }  
\end{equation*}
Note that $K$ is the size of the command string submitted when <enter> is pressed, and that a corner case must be accounted for, if multiple <enter> keystrokes are submitted in sequence, they will be prepended into the buffer of the next non-empty submission string, thereby imputing a decreased accuracy for the unnecessary keystroke events.

The aim of measure $A$ is to assess an operator state via accuracy for each strategic action issued. The value of the keystroke accuracy should nominally range in the interval $[0,1]$, the highest value is attained when the operator enters command actions correctly, while  lower values are attained when many modifications of the input buffer are observed.   

Note that all submissions end with an {<enter>} keystroke, which we do not count unless the buffer is empty (to prevent a division by zero).  When the user enters twice in a row the second enter will be seen as the first keystroke.

\begin{table}[h!]
    \centering
    \small
    \begin{tabular}{|c|c|a|a|a|}
        \hline 
        \rowcolor{LightGray}
         time & \multicolumn{3}{c|}{ user entered data } & stats \\ 
         \hline 
         date-time & raw &  {\bf normalized} & {\bf cmd}  & {\bf accuracy} \\
         \hline 
        & <Enter> & ssh && \\
         2023-02-23 & ssh s<BckSp> & root@ 	& ssh &  0.88  \\ 
         16:34:33 & root<LShft> &  10.10.111.102 & & \\   
         -0500  & @10.10.111.102 &   & & \\            
         \hline
         \end{tabular}
    \caption{Example of data collected, keystroke normalization, and statistical measures are shown for an example entry. 
    Data resulting from screen collection is given a white background, the inferred data is shown with gray background. 
    Keystroke normalization data, and command name (cmd) are interpreted from the raw entry as described while the accuracy statistics are calculated as a ratio of the length of normalized data to the length of raw input as the number of keystrokes.   }
    \label{tab:idf}
\end{table}

Table~\ref{tab:idf} illustrates the data processing with an example of raw input.  One can see the effects of applying rule $\mathcal{R}$ to normalize input, the command name is taken as the first white space separated string in the normalized keystroke column and appears as derived data in the column with heading {\bf cmd}.  This is used for indexing purposes and can further expedite the tagging tasks which are next described.  Additionally, the column labeled {\bf accuracy} calculates $A = \frac{23}{26}$ as the ratio of characters in the entered buffer to the total number of keystrokes.

\section{Labeling and Analysis}
\label{sec:labelingAndAnalysis}
Seeking to reason about the strategic actions of players, data such as that in Table~\ref{tab:idf} offers value, especially when command actions are reassembled and combined with statistics offering insight to player status, such as fatigue or processing speed levels.     
To further reason about strategy a type of data reduction is helpful. In particular, categorizing each command action in terms of a progression towards a goal offer a succinct and reduced description of an otherwise lengthy sequence of technical command action.  
One example of an ontological model which proposes a progression of actions for cyber operational commands is the MITRE ATT\&CK framework~\cite{mitre2023attack}. 
We therefore seek to tag our keystroke data with a simpler reduced description of a player's actions to derive sequences enabling a variety of insights about game strategy as well as player performance. 
However, the task entails expert tagging of the corpus of commands, and thus also demands design considerations.

To tag a corpus of user entered data we have designed and developed a new tool called {\it Pathfinder}\footnote{
https://github.com/xmohemx/HSM-} to efficiently and interactively label individual CTF command actions in terms of any hierarchical model.
Here, we describe its implementation and the operator effort required for the analysis to obtain an enriched dataset to include MITRE framework tags.

\subsection{{\it Pathfinder} Design and Implementation} 

\begin{figure*}[!htbp]
  \centering
  \includegraphics[width=0.95\textwidth]{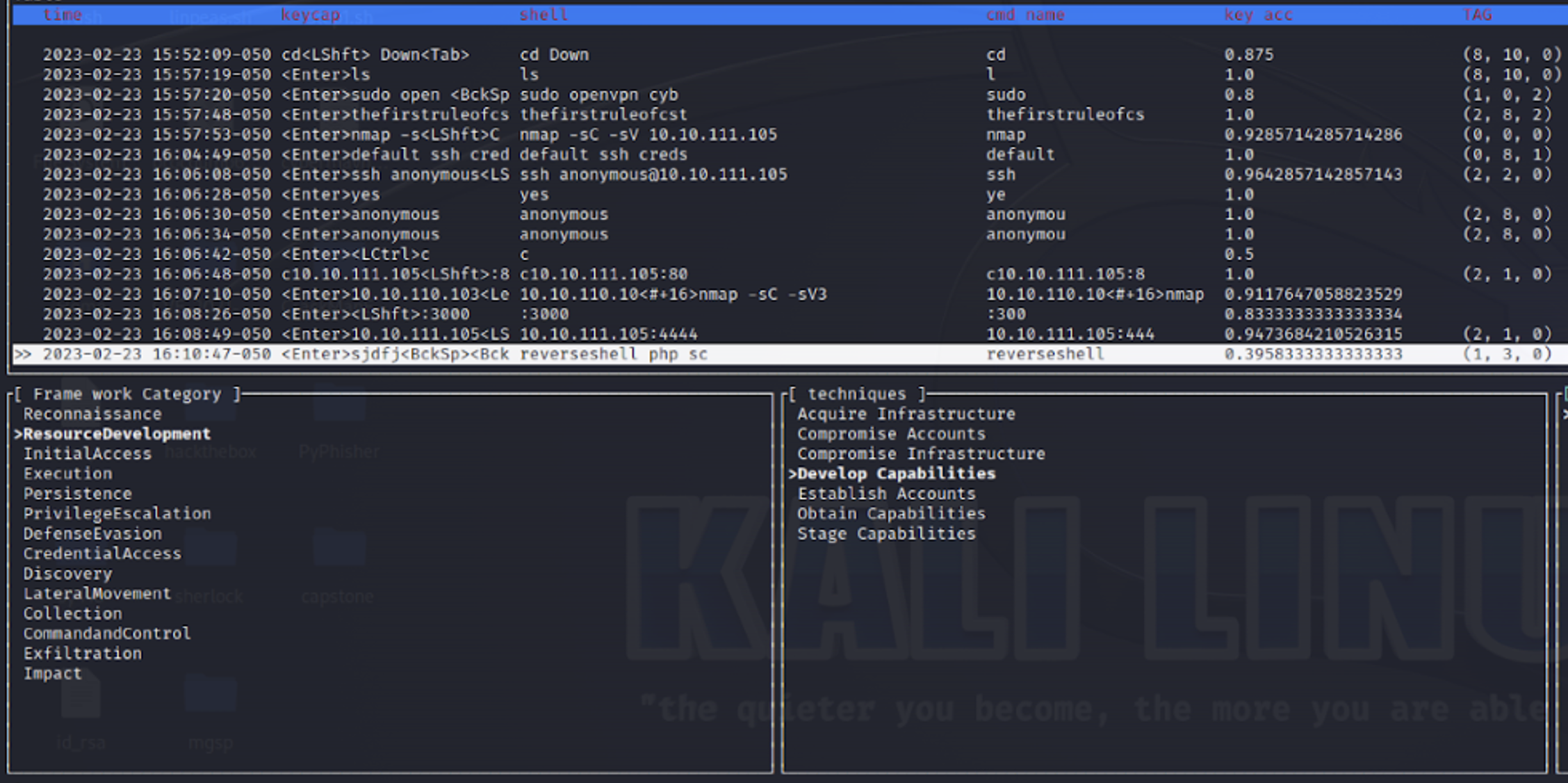}
  \caption{{\it Pathfinder} enables the human expert to tag or assign each action (by category, technique, sub-technique). {\it Pathfinder} makes this task more efficient and repeatable by presenting experts with a uniform interface and category browser speeding the expert's acclimation to categorical definitions. Further, the software provides shortcut keystrokes for navigation and tagging which aim to improve tagging accuracy.  }
  \label{fig:pathfinder}
\end{figure*}

Given the scale of data collected an efficient means to attribute labels to each command actions is sought.  The requirements are that {\it Pathfinder} can enable the efficient and systematized labeling of row oriented data into categories provided as a hierarchical data structure at initialization time.  As such, the ATT\&CK tactics and techniques can be configured as the tagging model to {\it Pathfinder} (Figure~\ref{fig:pathfinder}). 
Next, {\it Pathfinder} can be user to label comma separated value file data with the predefined set of labels. If the labels have a hierarchical data structure, {\it Pathfinder} will enumerate them into a coordinate system and allow the expert to tag rows of data with labels by using a dual navigation process.  To do this, {\it Pathfinder} presents the user with three views.  The first view is a spreadsheet view allowing the operator to view data for which a seek position can be moved using up/down arrows.  The second view navigates through a hierarchical structure of categories which define a seek position in ontology.  The ontology seek position is navigated as a tree using keystrokes on the keyboard.  The third view presents the full description of the category, identified by the seek position above. 
  Finally, the operator can bind the category to the data by pressing the space bar. 

\begin{figure}
    \centering
    \includegraphics[width=75mm]{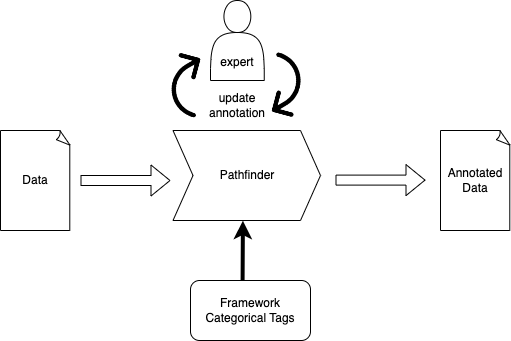}
    \caption{{\it Pathfinder} is an annotation tool configured with an ontological model which provides efficiencies for the expert in the tagging process.}
    \label{fig:my_label}
\end{figure}

\begin{figure*}[!htbp]
    \centering
    \includegraphics[width=0.95\textwidth]{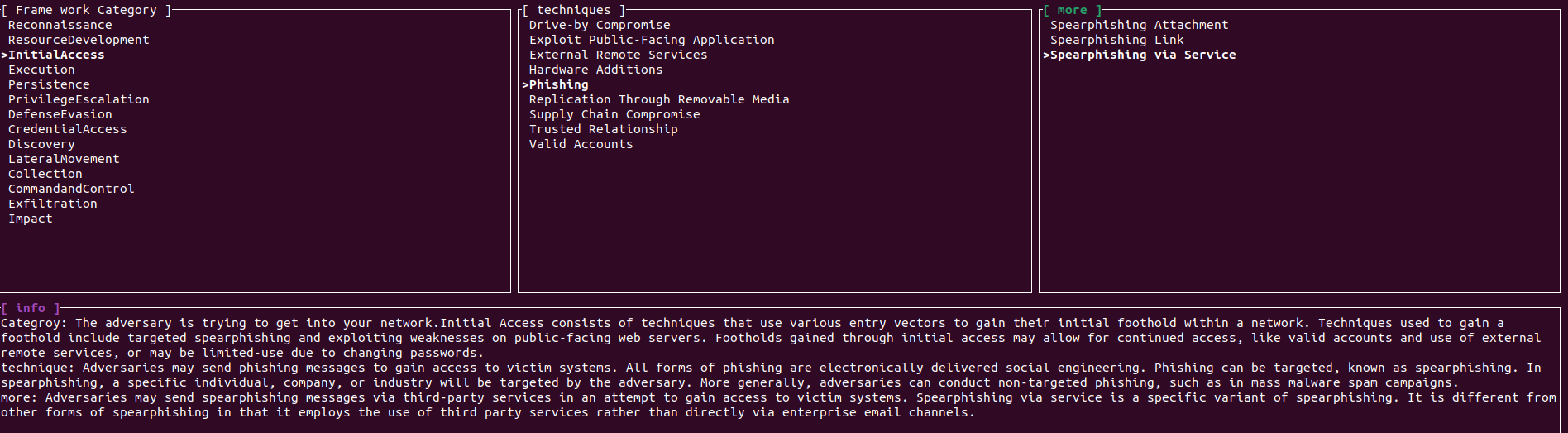}
    \caption{Ontology of MITRE ATT\&CK framework includes three levels: \underline{Categories} such as {\it Initial Access} expanded above; \underline{Techniques} within the category, such as {\it Phishing} expanded above; and \underline{Sub-Techniques} within the Technique, such as {\it spearfishing via services} selected above.  Assignment of the selected tag to a row of data will update the row to have the following attribution: $(2,4,2)$ which indicates the third (zero based index) category, the 5th technique (w/r/t the category), and the 3rd sub technique (w/r/t the technique).  Additionally, a definition of each level is supplied in the bottom view.
  }
    \label{fig:onto2}
\end{figure*}

{\bf Ontological model of command actions}.
By use of the official website for the MITRE ATT\&CK framework, we were able to parse and convert the hierarchical description of categories into an embedded python dictionary.
As of April 2023, the current version (v12) contains 14 Tactics (such as reconnaissance) and 193 Techniques (such as active scanning).
The hierarchy is a tree defining paths two to three links deep.  
The Path starts with a category, which divides into techniques, and further sub-divide into sub technical details.
We summarize the ontology as 14 categories, with hundreds of techniques, and sub-techniques.  An example taken from the ontology is shown in Figure~\ref{fig:onto2}.

The main task for {\it Pathfinder} is to enable the expert to label each action by assigning each row a (category, technique, sub-technique) designation (see Figure \ref{fig:pathfinder}).  The labeling process allows the user to navigate the spreadsheet rows (showing the time, keystroke log, and normalized command string as well as a score for keystroke accuracy) using arrow commands, and label hierarchy using letter keys, and allows the row to be attributed by the current label by entering return.  {\it Pathfinder} modifies data in-situation so the user's output is the modified comma separated value file, allowing the user to tag portions of the input file in a session of any length.  {\it Pathfinder} also offers definitions for each category so that a human operator can learn tag definitions if they are unfamiliar with them at the start, as shown in Figure~\ref{fig:onto2}.  

\subsection{Analysis}
One way to understand the data is how time is spent. This is possible at a granular level by looking at individual keystrokes and commands. A novice player may spend more time crafting individual commands and more time between commands than an experienced player. Time analysis is also possible at the ATT\&CK category, technique and sub-technique level. An aggressive attacker may spend little time and few commands on reconnaissance before launching initial access exploitation.

We also speculate that it is possible to fingerprint individual players based on a combination of their keystrokes, timing, and workflow. We do not yet know the minimum volume of data necessary to achieve a specific level of attribution accuracy and confidence.

Finally, the timing and attack flow of an individual is informative in real time when such data are available. For instance, backspace keystrokes and other typing errors that deviate from an individual's typical accuracy rate suggest the onset of fatigue and frustration.

\section{Case Study}
\label{sec:CaseStudy}
To exercise our data collection and analysis methodology, we conducted an exploratory experiment with 10 participants from one U.S. university.  Participants were volunteers from the university's CTF team. The experiment was approved by our Institutional Review Board (IRB).
The players participating in the experiment all had some experience with CTF activities prior to the experiment. To assess the level of experience, we asked participants: {\it How many years since you started doing CTF activities? (e.g., picoCTF, Hack The Box, etc.)}. Based on those responses, we grouped participants into an experience level of {\it Beginner} (0-1 years), {\it Intermediate,} (1-3 years) and {\it Advanced} (3+ years).

The CTF activity was an online game called {\it Cyber Mayhem} hosted by Hack The Box. This is a team game where teams must secure their systems, while also trying to exploit systems controlled by an opposing team.
In the game, participants were allowed to use any public resource to solve the problem including cheat sheets, exploit databases, and public code. Each player played the CTF game for 60 minutes. Components of the game included web exploitation, reverse engineering, and penetration testing. 

For the event itself, players selected their own teams of maximum size four and joined a Hack The Box (HTB) online game server.  Two volunteers therefore had to conduct non-team oriented learning modules during the same interval.  In our experiment, teams are organized as: Team 1 included players P1, P2, P5, and P8.  Team 2 includes players P3, P4, P6 and P7.  Players P9 and P10 were excluded from team play and conducted non-team related learning activities. 
All participants were given instructions on how to start the instrumented operating systems and then how to enter the HTB server game, in particular the experimenters needed to verify that collection was running prior to the game start.

At the completion of game (a duration of 60 minutes), teams receive a score which was based on the number of captured and retained flags leading to a score outcome of $3,950$ for the winning Team 2 and $3,146$ for Team 1.  
We summarize the resulting scores in Table~\ref{tab:stats}.   
During game play the experimenters were on hand to observe and verify collection integrity, and following the game, experimenters gathered all data files using Cluster SSH to a central repository for subsequent analysis. 

During the game, we observed lively team play and dynamic actions which resulted in brief in person discussions.  Players would regularly check the scoreboard, made available at the HTB game server, along with launching many commands and using web browsers as search tools for a variety of different types of information. 

After the conclusion of the game, players informally discussed among themselves strategies and particularly impactful game actions. 

\begin{table}[t]
\renewcommand{\arraystretch}{1}
\setlength{\tabcolsep}{4pt}
\selectfont\centering

{\small
\begin{tabular}{l c c c c c}
\toprule
\textbf{Player}	&
\textbf{Team}	&
\textbf{Gender} &
\textbf{CTF Exp.$^{1}$}	&
\textbf{data} &
\textbf{cmd Accuracy} \\
& & & & cmds video & [mean] <std> \\
\midrule

P1	&	1 & F     &	I	&	62 426M	&	[0.7646] <0.2981>  \\
P2	&	1 & M     &	B	&	348 381M	&	[0.5264] <0.4303>  \\
P3	&	2 & M     &  B	&	232 651M	&	[0.5531] <0.4035> \\
P4	&	2 & F     &  I	&	306 351M	&	[0.8009] <03342>  \\
P5	&	1 & F     &  B	&	279 325M	&	[0.7604] <0.2750>  \\
P6	&	2 & M     &  B	&	254 425M	&	[0.7597] <0.3898>   \\
P7	&	2 & M     &  A	&	461 461M	&	[0.7080] <0.3461>  \\
P8	&	1 & M     &  I	&	364 432M	&	[0.6713] <0.3482>  \\
P9	&	- & M     &  B	&	316 400M	&	[0.6761] <0.4398>  \\
P10	&	- & M  & I	&	372 1196M	&	[0.6880] <0.3404>  \\

\bottomrule
\multicolumn{6}{l}{\footnotesize{$^1$B: Beginner (0-1 years), I: Intermediate (1-3 years), A: Advanced (3+ years) }}\\
\end{tabular}
}

\caption{
Summary of case study participants with team, gender, CTF experience, data collection (number of commands, video capture size), accuracy (mean and standard deviation) of collected commands.
}
\label{table:studysummary}
\end{table}

\subsection{Findings from Player Accuracy}
One way to understand accuracy is in relationship to expertise. Each of our participants performed an average of 299 commands during the exercise with an average keystroke 
 accuracy of 69.1\%.
The winning team, Team 1, had 19\% fewer combined commands (1,053) but 4\% lower keystroke accuracy (68.1\%) compared with Team 2 (1,253 combined commands and 70.5\% keystroke accuracy).

We created a model based on the accuracy of each player's inputted commands. In this model, 10 distinct players compete against the clock to attack each other in the HTB environment, and based on the data from their inputs, we computed the confidence accuracy. 
Among our participants, P1 had the fewest commands (62) and the lowest mean accuracy (52.6\%). This player also had beginner level CTF experience. The highest keystroke accuracy was from P4 (80.1\%) who had intermediate CTF experience. Using this example, it is simple to distinguish between novice players with some cyber experience and experienced players All of these players were subjected to the same environment and time limitations, and by providing a model that illustrates these accuracies. 
After game play, we applied the parsing techniques defined earlier by equation 
${\mathcal{R}}(x)$. 
\begin{table}[h!]
    \centering
    \small
    \begin{tabular}{l r r r}
\toprule
\textbf{Grouping} &
\textbf{Commands} & 
\multicolumn{2}{c}{\textbf{Performance}} \\
& \textbf{mean number} & \textbf{mean accuracy} & \textbf{mean score}  \\
\midrule 
\multicolumn{4}{l}{\textbf{By Team}} \\
 Team 1 & 239.75 & 0.6807 & \textbf{3,950} \\ 
 Team 2 & \textbf{336.50} & \textbf{0.7054} & 3,146  \\ 
\hline 
\multicolumn{4}{l}{\textbf{By Experience}} \\
Beginner & 247.2 & 0.6552 & 3,548\\
Intermediate & 324 & 0.7350 & \textbf{3,682}\\
Advanced & \textbf{461} & \textbf{0.7625} & 3,146\\
\bottomrule

    \end{tabular}
    \caption{Accuracy and score by team and by CTF experience.}
    \label{tab:stats}
\end{table}

Table~\ref{tab:stats} provides a summary of the mean number of commands, accuracy, and game score for each of the two teams and by experience level.  Our observations indicate that the number of commands issued and keystroke accuracy seem to correspond with player experience.  On the other hand, score outcomes, at least in this case, appear to be elusive of obvious trends.  
Our data seems to suggest that: 
\begin{itemize}
    \item Players with more experience may have access to a larger repertoire of command actions and issue greater levels of commands.
    \item Experience and/or practice may correspond to greater keystroke accuracy, even when more command actions are performed.
\end{itemize}  
Said differently, learners who are developing their skills may be gaining more commands within their repertoire and improving their focus and control in times of stress and uncertainty.

One may hypothesize that large command repertoire and high keystroke accuracy would also correlate with higher performance scores if they are indicative of focus, attention, and expertise; on the contrary, we observe
that team score may be unrelated or entail other factors. 
Not surprisingly, and similarly to other team sports, this indicates that team play and outcome score may engender a more complex dynamic then individual player experience alone.  
Note that while Team 2 has 0.0247 greater keystroke accuracy than Team 1, their score is 20\% less from that of Team 1. 

Team 1 is comprised of two beginner and two intermediate player (i.e., BBII) while Team 2 is comprised of two beginner, one intermediate, and one advanced player (i.e., BBIA).  We posit that team dynamics may be just as important as individual players.  The effects of team play may include the possibility that complex synergies can emerge; in our example, two intermediate players produced a higher score outcome than an intermediate and an advanced player.  
Different variations of team configurations and their effect on performance is a largely unstudied area in CTF games, but could potentially be measured using our methodology.  

\begin{figure}
    \centering
    \includegraphics[width=1.0\columnwidth]{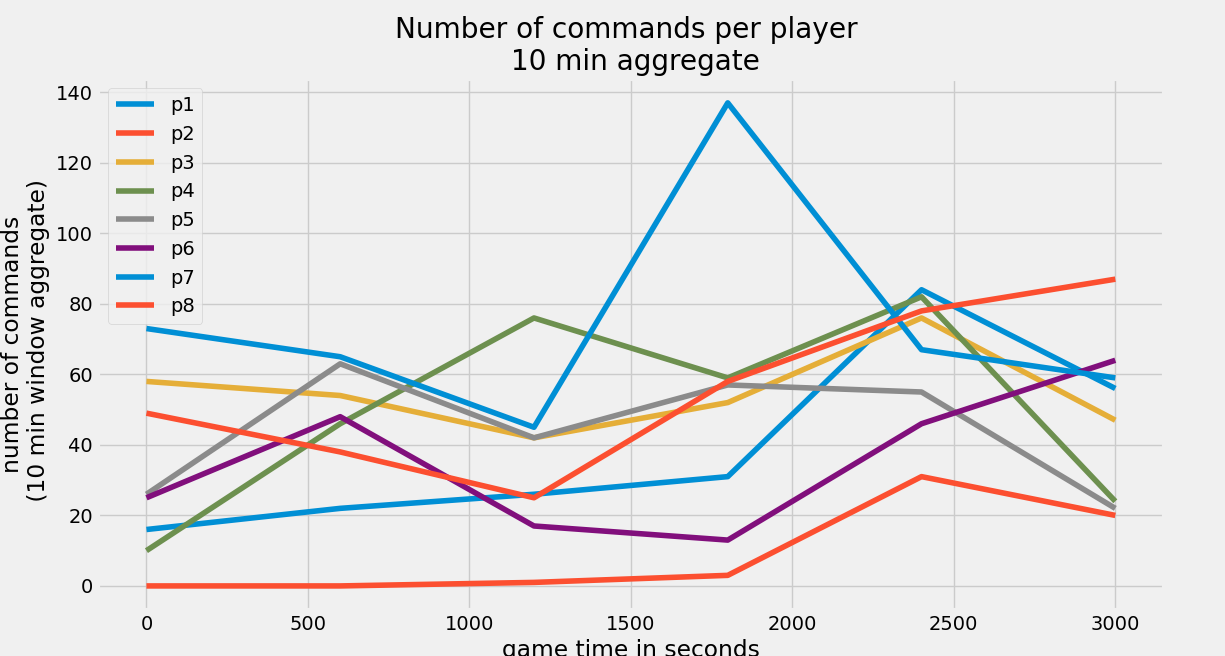}
    \caption{Command counts time series per player where counts are aggregated into 10 minute non-overlapping windows during a one-hour game.}
    \label{fig:counts}
\end{figure}

\begin{figure}[!htbp]
\centering
    \includegraphics[width=1.0\columnwidth]{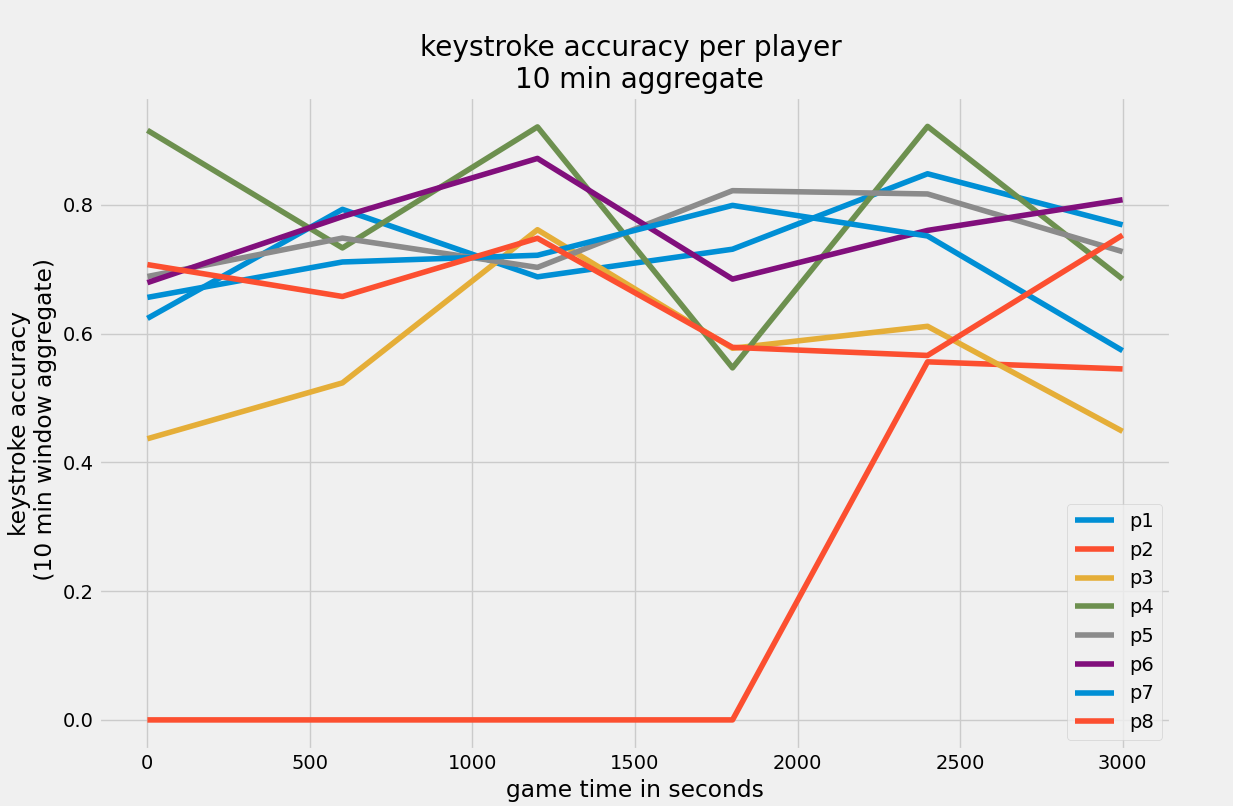}
    \caption{Keystroke accuracy time series per player where counts are aggregated into 10-minute non-overlapping windows during a one-hour game.  Note that accuracy measures zero during periods of inactivity.}
    \label{fig:accuracyOverTime}
\end{figure}

Finally, we calculated the {\it number of command actions}  and {\it keystroke accuracy} for each participant in 10 minute non-overlapping intervals over the course of the hour game (see Figure~\ref{fig:counts} and Figure~\ref{fig:accuracyOverTime}). 
Fatigue, in general, is often a consideration in strenuous team play, and usually follows a particularly active or demanding portion of play; as such, it may be possible to observe in data.  
While accuracy varies over time, we cannot say with certainty whether or not these changes are caused by stress or fatigue.  Note that in Figure~\ref{fig:accuracyOverTime} only P4 and P8 drop accuracy after the first 10 minutes, and only P6 and P8 increase accuracy in the last 10 minutes; all other players increase accuracy in the first 10 minutes and drop in accuracy in the last 10 minutes.  
Despite our small sample size, it may be interesting to test for a pattern of {\it warm-up} and {\it cool-down} as suggested by these limited observations.  

These insights represent potentially relevant and notable findings for different audiences and use cases. A deeper and more systematic study using these techniques may provide necessary validation. We therefore plan to carry this work forward as future work to include larger observations capable of generating data sets to address or confirm some of these possibilities. 

\subsection{Findings from ATT\&CK Labeling}

We observed a total of 2,994 commands from our 10 participants. After normalizing the command data and labeling commands using {\it Pathfinder}, we produced a starting point for a statistical study of command action sequences as categorized by ontologies such as MITRE ATT\&CK framework. Although we have not yet completed labeling every command, this dataset remains promising.

In our preliminary set of labeled data, we have observed a view of each player's behaviors in correspondence with the ATT\&CK framework. Each human participant plays the game slightly differently, spending their time and attention in a unique order and quantity to successfully accomplish their tasks.
In the future, we hope to further analyze patterns across the players including how much time they spend in each phase of the ATT\&CK framework and how individual players move between phases.

\section{Conclusion}
\label{sec:Conclusion}
We have presented a methodology for collecting, labeling, analyzing, and operationalizing insights from human participants in capture-the-flag games. Using a new labeling tool, we applied the MITRE ATT\&CK framework to create a data set that shows value in creating attack flows.  

There were several limitations to our study. The number of participants was deliberately small and diverse (background experience), which we thought would be best to test the methodology.  We did not collect a baseline of normal keystroke accuracy to compare individuals against their typical error rates. Another limitation is that we did not collect players' intent as they played the game. This could have been done with a think aloud protocol or interviews, and would have enabled us to validate our ATT\&CK labels. One means to improve the analysis of participants would be to make a fingerprint based on their command usage, error patterns, time spent researching, or movements within the network once exploitation is successful, enabling attribution to that participant or team.

There are rich opportunities to combine our work with other recent advances. For instance, we could apply the work of Kim et al. to score the attacks used by players in the CTF~\cite{kim2021cyber}. We could also use the techniques of Al-Shaer, Spring, and Christou to group actions into fine-grain and coarse-grain associations with which to predict attackers' next behavior using a model built on imperfect quantities of past behavior~\cite{al2020learning}.

We intend to extend this line of research in several dimensions. First, to compliment ATT\&CK, MITRE also created and maintains a public repository of defender tactics and techniques called D3FEND~\cite{mitre2023defend}. Although we had some defensive data from the CTF, we did not attempt to do MITRE D3FEND labeling but defer that for future work. We plan to instrument {\it Pathfinder} to make this possible, as well as further develop tools to make data tagging more efficient.  Our methodology enables a type of team play analysis, where team performance outcomes can be examined when teams are comprised of various skill levels.  We plan to expand the study to explore the relation between keystroke accuracy, learner experience, and score outcome, as well as team composition. Next, we aim to examine performance characteristics, such as what human players do when they get stuck, tired, or frustrated. Finally, we have begun to experiment with artificial intelligence that could emulate the behavior of a particular player and mimic their attack narratives, or suggest new ones, such as move options to augment a human player. We also seek to understand the privacy and security implications of such attributions.

\begin{acks}
We thank the anonymous reviewers for their comments and suggestions. 
The views and conclusions expressed in this paper are those of the authors, and do not necessarily represent those of the Department of Defense or U.S. Federal Government.
\end{acks}

\bibliographystyle{ACM-Reference-Format}
\bibliography{bibliography}

\end{document}